\documentclass[11pt]{llncs}
\usepackage{fullpage}

\newcommand{\Oh}[1]
    {\ensuremath{\mathcal{O}\!\left( {#1} \right)}}
\newcommand{\occ}
    {\ensuremath{\mathrm{occ}}}
\newcommand{\rank}
    {\ensuremath{\mathrm{rank}}}
    
\begin{document}

\title{Grammar-Based Construction of Indexes\\for Binary Jumbled Pattern Matching}
\author{Travis Gagie}
\institute{Department of Computer Science and Engineering\\Aalto University}
\maketitle

\begin{abstract}
We show how, given a straight-line program with $g$ rules for a binary string $B$ of length $n$, in $\Oh{g^{2 / 3} n^{4 / 3}}$ time we can build a \((2 n H_0 (B) + o (n))\)-bit index such that, given $m$ and $c$, in $\Oh{1}$ time we can determine whether there is a substring of $B$ with length $m$ containing exactly $c$ copies of 1.  If we use $\Oh{n \log n}$ bits for the index, then we can list all such substrings using $\Oh{m}$ time per substring.
\end{abstract}

\section{Introduction} \label{sec:introduction}

Motivated by problems in mass spectrometry, several researchers have studied the problem of jumbled pattern matching, i.e., finding substrings of a given text that consist of a given multiset of characters.  Building indexes for jumbled pattern matching has turned out to be a challenging problem: currently, we can build reasonably-sized indexes only for binary strings and, even then, the construction time is superlinear and the indexes report only whether there exists a matching substring.

Cicalese, Fici and Lipt\'ak~\cite{CFL09} showed how, given a binary string \(B [1..n]\), in $\Oh{n^2}$ time we can build a linear-space index such that, given $m$ and $c$, in $\Oh{1}$ time we can determine whether there is a substring of $B$ with length $m$ containing exactly $c$ copies of 1.  Their key observation was that, if one substring of length $m$ contains fewer than $c$ copies of 1 and another contains more, then a third such substring must contain exactly $c$ copies.  Their index is a table $T$ saying, for \(1 \leq m \leq n\), how many and how few 1s there can be in a substring of length $m$.

Burcsi, Cicalese, Fici and Lipt\'ak~\cite{BCFL10} (see also~\cite{BCFL12a,BCFL12b}) and Moosa and Rahman~\cite{MR10} independently reduced Cicalese, Fici and Lipt\'ak's construction time to $\Oh{n^2 / \log n}$, then Moosa and Rahman~\cite{MR12} reduced it to $\Oh{n^2 / \log^2 n}$ in the RAM model.  Cicalese, Laber, Weimann and Yuster~\cite{CLWY12} showed how to build in $\Oh{n^{1 + \epsilon}}$ time an approximate index, which may return false positives when the query is close (in a sense depending on $\epsilon$) to one for which there is a matching substring.  Badkobeh, Fici, Kroon and Lipt\'ak~\cite{BFKL12} showed how to build in $\Oh{r^2 \log r}$ time a potentially smaller index that answers queries in $\Oh{\log n}$ time, where $r$ is the number of runs in $B$ (i.e., maximal substrings containing only 0s or only 1s).  Recently, Giaquinta and Grabowski~\cite{GG12} gave several time-space tradeoffs based on Badkobeh et al.'s construction.

Suppose that, as we receive $B$ as input, we store it as a compressed bitvector that takes \(\Oh{n H_0 (B)} + o (n)\) bits, where \(H_0 (B) \leq 1\) is the 0th-order empirical entropy of $B$.  With this bitvector, we can answer rank queries on $B$ in $\Oh{1}$ time; given a position, a rank query returns the number of 1s in the prefix ending at that position.  We can loop through all $\Oh{n^2}$ substrings of $B$ in increasing order of length and count the number of 1s in each substring using two rank queries.  This way, in $\Oh{n^2}$ time we can compute $T$ row by row using \(\Oh{n H_0 (B)} + o (n)\) bits of workspace.

Fici and Lipt\'ak~\cite{FL11} (see also~\cite{BFKL12}) observed that, if we increase $m$ by 1, then the minimum number of 1s can only stay the same or increment, as can the maximum number of 1s.  Therefore, we can encode $T$ as two $n$-bit binary strings $B_{\min}$ and $B_{\max}$ in which 1s indicate increments.  Given $m$ and $c$, in $\Oh{1}$ time we can determine whether there is a substring of $B$ with length $m$ containing exactly $c$ copies of 1, by checking whether \(B_{\min}.\rank (m) \leq c \leq B_{\max}.\rank (m)\).  We now note that \(B_{\min}.\rank (n) = B_{\max}.\rank (n) = B.\rank (n)\), so \(H_0 (B_{\min}) = H_0 (B_{\max}) = H_0 (B)\).  Therefore, if we store $B_{\min}$ and $B_{\max}$ as compressed bitvectors, then they take a total of \(2 n H_0 (B) + o (n)\) bits and support rank queries in $\Oh{1}$ time.  We can build these compressed bitvectors incrementally as we compute $T$ row by row, so we still use \(\Oh{n H_0 (B)} + o (n)\) bits of workspace overall.

In this paper we show how, given a straight-line program (SLP) for $B$ with $g$ rules (i.e., a context-free grammar with $g$ rules in Chomsky normal form that generates $B$ and only $B$), we can build Cicalese, Fici and Lipt\'ak's table $T$ in $\Oh{g^{2 / 3} n^{4 / 3}}$ time.  Informally, a string has a small SLP if LZ77 compression works well on it, and vice versa~\cite{Ryt03}; by the analysis of LZ78, we can assume \(g = \Oh{n / \log n}\) so our time bound is $\Oh{n^2 / \log^{2 / 3} n}$ even when $B$ is not compressible.  We also show how, by indexing substrings and using a total of $\Oh{n \log n}$ bits, we can list all substrings of length $m$ containing exactly $c$ copies of 1 using $\Oh{m}$ time per substring.  In the full version of this paper we will improve our construction time slightly by combining our approach with Moosa and Rahman's; we will also show that our construction still takes \(\Oh{n H_0 (B)} + o (n)\) bits of workspace.

\section{Construction} \label{sec:construction}

Suppose we have a SLP $S$ for $B$ with $g$ rules.  Gawrychowski~\cite{Gaw12} recently showed how, for any $\ell$, in $\Oh{g + n / \ell}$ time we can turn $S$ into an SLP $S'$ with $\Oh{g}$ rules such that all new nonterminals expand into strings of length at most $\ell$ and $B$ can be written as the concatenation of $\Oh{n / \ell}$ new nonterminals' expansions.  It follows that, setting \(\ell = (n / g)^{2 / 3}\), in \(\Oh{g + g^{2 / 3} n^{1 / 3}} = \Oh{n}\) time we can split $B$ into $\Oh{g^{2 / 3} n^{1 / 3}}$ blocks \(B_1, \ldots, B_b\) of length at most \((n / g)^{2 / 3}\) such that there are $\Oh{g}$ distinct blocks \(\mathbf{B_1}, \ldots, \mathbf{B_d}\).

\begin{lemma} \label{lem:blocks}
In $\Oh{n}$ time we can split $B$ into $\Oh{g^{2 / 3} n^{1 / 3}}$ blocks of length at most \((n / g)^{2 / 3}\) such that there are $\Oh{g}$ distinct blocks.
\end{lemma}

For each distinct block $\mathbf{B_i}$, \(1 \leq i \leq b\), we build a table $\mathbf{T_i}$ saying, for each length $m$ at most the length of $\mathbf{B_i}$, how many and how few 1s there can be in a substring of $\mathbf{B_i}$ with length $m$.  This takes a total of $\Oh{n^{4 / 3} / g^{1 / 3}}$ time.  For each possible pair of distinct blocks $\mathbf{B_i}$ and $\mathbf{B_j}$, \(1 \leq i, j \leq b\), we build a table $\mathbf{T_{i, j}}$ saying, for each length $m$ at most the length of their concatenation \(\mathbf{B_i B_j}\), how many and how few 1s there can be in a substring of \(\mathbf{B_i B_j}\) with length $m$ that starts in $\mathbf{B_i}$ and ends in $\mathbf{B_j}$.  This takes a total of $\Oh{g^{2 / 3} n^{4 / 3}}$ time.

For \(1 \leq i \leq j \leq b\), we build a table $T_{i, j}$ saying, for each length $m$ at most the length of the concatenation \(B_i \cdots B_j\), how many and how few 1s there can be in a substring of \(B_i \cdots B_j\) with length $m$ that starts in $B_i$ and ends in $B_j$.  If \(j = i\) then $T_{i, j}$ is the table for the distinct block of which $B_i$ is a copy.  If \(j = i + 1\) then $T_{i, j}$ is the table for the pair of distinct blocks of which $B_i$ and $B_j$ are copies.  Only when \(j > i + 1\) do we need to build a new table.

Let $m_{i + 1, j - 1}$ be the length of \(B_{i + 1} \cdots B_{j - 1}\) and let $c_{i + 1, j - 1}$ be the number of 1s in \(B_{i + 1} \cdots B_{j - 1}\).  Let $B'$ be a substring of \(B_i \cdots B_j\) and let $B''$ be the substring of \(B_i B_j\) obtained by removing \(B_{i + 1} \cdots B_{j - 1}\) from $B'$.  Notice that the length of $B'$ is the length of $B''$ plus $m_{i + 1, j - 1}$, and the number of 1s in $B'$ is the number of 1s in $B''$ plus $c_{i + 1, j - 1}$.  It follows that we can build the table $T_{i, j}$ by copying the table for the pair of distinct blocks of which $B_i$ and $B_j$ are copies, adding $m_{i + 1, j - 1}$ to all the lengths and adding $c_{i + 1, j - 1}$ to all counts of 1s.  Building all such tables takes a total of $\Oh{g^{2 / 3} n^{3 / 4}}$ time.  Merging them to form $T$ also takes $\Oh{g^{2 / 3} n^{3 / 4}}$ time.

\begin{lemma} \label{lem:construction}
We can build $T$ in $\Oh{g^{2 / 3} n^{3 / 4}}$ time.
\end{lemma}

This lemma immediately implies the following theorem.  In the full version of this paper, we will improve these results slightly by using Moosa and Rahman's result to build the tables for the blocks and pairs of blocks.

\begin{theorem} \label{thm:construction}
Given an SLP with $g$ rules for \(B [1..n]\), in $\Oh{g^{2 / 3} n^{4 / 3}}$ time we can build a \((2 n H_0 (B) + o(n))\)-bit index such that, given $m$ and $c$, in $\Oh{1}$ time we can determine whether there is a substring of $B$ with length $m$ containing exactly $c$ copies of 1.
\end{theorem}

\section{Listing} \label{sec:listing}

If we are willing to increase our space usage by a logarithmic factor, then it is not difficult to turn indexes for detecting jumbled pattern matches into indexes for locating them.  For the sake of simplicity, assume $n$ is a power of 2.  For \(1 \leq i \leq \log n - 1\) and \(0 \leq j \leq 2^i - 2\), we build an index for detecting jumbled pattern matches in \(B [j n / 2^i + 1..(j + 2) n / 2^i]\).  That is, we build indexes for \(B [1..n], B [1..n / 2], B [n / 4 + 1..3 n / 4], B [n / 2 + 1..n], B [1..n / 4], B [n / 8 + 1..3 n / 8], B [n / 4 + 1..n / 2], B [3 n / 8 + 1..5 n / 8], B [n / 2 + 1..3 n / 4], \ldots\).  These indexes take a total of $\Oh{n \log n}$ bits.

We can visualize this as \(\log n - 1\) sets of overlapping segments: the first set consists of the segment \(B [1..n]\), i.e., the whole string; for \(2 \leq i \leq n\), the $i$th set consists of two layers of disjoint segments of length \(n / 2^{i - 1}\), with the first layer starting at position 1 and the second layer starting at position \(n / 2^i + 1\).  (We use ``segment'' instead of ``block'' to avoid confusion with Section~\ref{sec:construction}.)  Notice that, for any substring of length $m$ and any set whose segments have length at least \(2 m\), that substring is completely included in at least 1 segment and at most 2 segments in that set.

Given $m$ and $c$, we first query the index for the whole string \(B [1..n]\).  For \(2 \leq i \leq \lfloor \log m \rfloor - 1\), if we found a segment $B'$ in the \((i - 1)\)st set that contains a match, then we query the indexes for the segments in the $i$th set that are completely contained in $B'$.  Let \occ\ be the number of jumbled pattern matches for $m$ and $c$.  We make $\Oh{\occ \log n}$ queries to indexes for segments, which takes $\Oh{\occ \log n}$ time, and find $\Oh{\occ}$ segments of size at most \(2 m\) that together completely include all the jumbled pattern matches for $m$ and $c$.  We then scan those segments and find all the matches, which takes $\Oh{\occ (m + \log n)}$ time.

We can reduce our time bound to $\Oh{\occ\,m}$ by using a different strategy when \(m < \log n\).  For \(1 \leq m < \log n\) and \(0 \leq c \leq m\), we store a bitvector with 1s indicating which segments in the \((\lfloor \log m \rfloor - 1)\)st set contain a substring of length $m$ with $c$ 1s.  Each bitvector takes $\Oh{n / m}$ space, so we use $\Oh{n \log n}$ bits in total.  Given \(m < \log n\) and $c$, we use $\Oh{1}$-time select queries on the appropriate bitvector to find the segments to scan.

Straightforward calculation shows that, if we index each segment in time quadratic in its length, then we use $\Oh{n^2}$ time overall.  To index all the segments in $\Oh{g^{2 / 3} n^{4 / 3}}$ time, for each segment longer than $g$, we build an SLP for that segment with $\Oh{n'}$ rules, where $n'$ is the length of the segment.  If \(n' > g\), then we do this by restricting the original SLP to generate only the segment, which introduces $\Oh{g}$ new nonterminals.  If \(n' \leq g\), then we build the new SLP from scratch.  Once we have the new SLP, by Theorem~\ref{thm:construction}, we can build the table for it in $\Oh{g^{2 / 3} (n')^{4 / 3}}$ time.  Summing up over all of the segments, we use a total of $\Oh{g^{2 / 3} n^{4 / 3}}$ time.

\begin{theorem} \label{thm:listing}
Given an SLP with $g$ rules for \(B [1..n]\), in $\Oh{g^{2 / 3} n^{4 / 3}}$ time we can build an $\Oh{n \log n}$-bit index such that, given $m$ and $c$, we can list all substrings of length $m$ containing exactly $c$ copies of 1 using $\Oh{m}$ time per substring.
\end{theorem}

\section{Future Work} \label{sec:future}

As noted before, in the full version of this paper we will improve Lemma~\ref{lem:construction} and Theorem~\ref{thm:construction} (and, thus, Theorem~\ref{thm:listing}) slightly by using Moosa and Rahman's result to build the tables for the blocks and pairs of blocks.  We will also show that our construction takes \(\Oh{n H_0 (B)} + o (n)\) bits of workspace.  Finally, we will also consider some new problems.  For example, suppose we have a tree whose nodes are labelled with characters from a constant-sized alphabet.  If we build a perfect hash table for the Parikh vectors of all the subtrees, then we use $\Oh{n}$ space (measured in words) and later we can list the subtrees with a given Parikh vector using $\Oh{1}$ time per subtree.  This approach can be applied to jumbled pattern matching in strings, as well, but in general the space usage increases to $\Oh{n^2}$ (or to $\Oh{n^{2 d}}$ for $d$-dimensional arrays).

\section*{Acknowledgments}

Many thanks to Ferdinando Cicalese, Emanuele Giaquinta, Szymon Grabowski, Kalle Karhu, Zsuzsanna Lipt\'ak, Simon Puglisi and Jorma Tarhio, for helpful comments.

\bibliographystyle{plain}
\bibliography{jumbled}

\end{document}